# Survey on *Essential and Accidental Real-Time Issues* in Software Engineering

**Furrakh Shahzad[1], Dr. Maruf Pasha[2], Dr. Urooj Pasha[3], Bushra Majeed[2], Khurram Shahzad[2]**

[1]Department of Computer Science, Pakistan Institute of Engineering and Technology, Multan, Pakistan
[2]Department of Information Technology, Bahauddin Zakariya University, Multan, Pakistan
[3]Institute of Management Sciences, Bahauddin Zakariya University, Multan, Pakistan
Email: farrukhshahzad@piet.edu.pk, maruf.pasha@bzu.edu.pk, urooj.pasha@bzu.edu.pk, engr.bushra@yahoo.com, khurram0732@hotmail.com

## Abstract

Software product lines have recently been presented as one of the best promising improvements for the efficient software development. Different research works contribute supportive parameters and negotiations regarding the problems of producing a perfect software scheme. Traditional approaches or recycling software are not effective to solve the problems concerning software competence. Since fast developments with software engineering in the past few years, studies show that some approaches are getting extensive attention in both industries and universities. This method is categorized as the software product line improvement; that supports reusing of software in big organizations. Different industries are adopting product lines to enhance efficiency and reduce operational expenses by way of emerging product developments. This research paper is formed to offer in-depth study regarding the software engineering issues such as complexity, conformity, changeability, invisibility, time constraints, budget constraints, and security. We have conducted various research surveys by visiting different professional software development organizations and took feedback from the professional software engineers to analyze the real-time problems that they are facing during the development process of software systems. Survey results proved that complexity is a most occurring issue that most software developers face while developing software applications. Moreover, invisibility is the problem that rarely happens according to the survey.

## 1. Introduction

Essential properties are the properties considered as the core thing: An engine, four wheels, steering and a transmission are crucial to make a complete car. These are core features. A car could and could not have a V6 or a V8, studded snow wheels or racing slicks, a manual or an automatic transmission. These are the "accidental" or non-functional properties [1]. Software engineering usually includes the clarification and identification of system necessities, understanding and organizing the problem world, the arrangement, and requirement of the hardware and software machine that can guarantee the satisfaction of needs in problem world [2]. The field of software engineering has seen an explosion in curiosity in last few years. The industrial organizations have utilized product lines over the persistent period to uplift their productivity and reduce the operation expenses by way of emerging several product commonalities [3].

Till 2025, ever-increasing interests will be put on the computer based software to deliver safe, reliable and secure information technology, to offer new products in the market. Moreover, to support the management of multi-cultural worldwide enterprises, to allow rapid revision to change and to aid people to cope with the complex masses of information and facts, major changes are needed to be made in software systems. Such demands will cause major alterations in the processes presently used to describe, design, develop, arrange, and evolve a diverse range of software-intensive schemes [4] [5].

Software engineering is basically an organized and systematic method of the development, process, maintenance, and withdrawal of software. There are some important problems that software engineering usually faces.

**The Problem of Scale:** An important problem in the software engineering is a problem of scale; enlargement of large system necessitates a diverse set of approaches as compared to developing small application or system. In simple words, the approaches used for the developing of the small applications normally do not be compatible with the large systems. Different varieties of approaches have to be used for engineering the big software systems.

**Cost, Quality, and Schedule:** Cost of engineering software is basically a cost of the set of resources used in the development process of this system. For the software, these resources include hardware, software, the manpower, electricity and the other provision resources. Mostly, the manpower is a predominant component, as software engineering is mainly labor-intensive and the charge of the computing systems is getting quite low.

The quality of software systems is also very important because users distinguish the same product of two different brands regarding price and quality. If the quality of the software is low, then its market credibility will be reduced. We can see eminence of a software product as having three scopes: Product Transition, Product Operation, and Product Revision. Scheduling of every step is very important. Without proper scheduling of all phases of software development, software cannot be built on time having all functionalities with full accuracy.

**Consistency Problem:** Though low cost, high quality, great cycle time sche-duling are the main points of any scheme, for organization, consistency is also very demanding. An organization simply does not just need high quality and low-cost software, but it wants all these things consistently.

The purpose of this research work is to identify essential and accidental difficulties that professional software engineers face while working. Paper also describes the level and scope of these problems. Survey also tells that about the projection of these problems, occurring chances of such issues and efforts are required to handle such issues.

## 2. Literature Review

The impression that software has to be "engineered" induces an image of care, assurance, and rigor. In the 1980s, professional courses and university education emerged with the "software engineering" as a crucial component of their label and content; and these courses seemed to offer a more realistic understanding of the computer expertise than did more traditional Information technology or computer science courses [6]. The process of software engineering is the key to the development of pleasing software. Its design demands development and study of process models [7].

The software is progressively becoming the greatest success aspect for the future products (aircraft, radios, and automobiles) and services (defense, financial and communications,). It delivers both, competitive diversity and quick adaptability to modest change [8]. Software companies habitually start with a particular idea—and thus with a new product. With the passage of time, if the company goes up then product matures, and organization understands that it can practice the same idea (or little variations of it) to advance a set of the goods [9]. Some issues that cause the failure of software projects are mention in the survey result [10] shown in Figure 1.

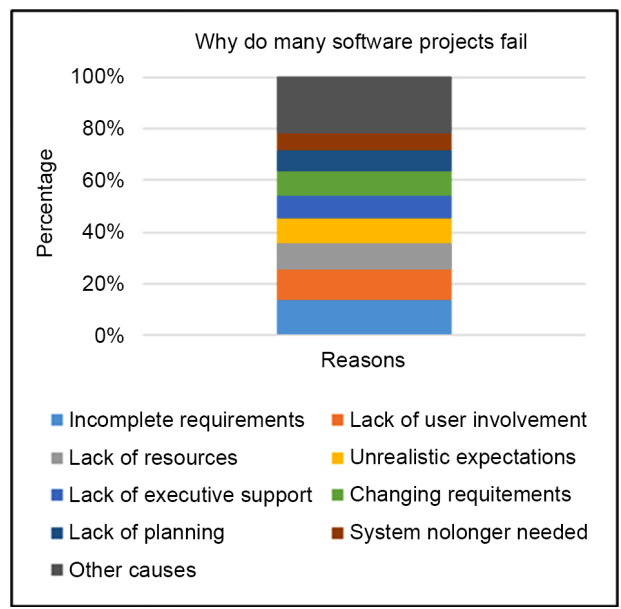

**Figure 1.** Study by Standish group involving 350 companies from 1994/95.

Like other engineers, a software engineer must be skillful in following points [11]:

- Theoretical fundamentals of discipline;
- Designing methods of discipline;
- Tools and technology of discipline;
- He/she must be talented to keep his/her understanding of the current as well as new techniques and approaches;
- Interaction with professionals;
- Understand, formalize, model and analyze the latest problems;
- Recognize a periodic problem, and reprocess or adapt acknowledged solutions;
- Manage a procedure and organize the work of different persons.

Small and Medium Enterprises (SMEs) have some specific issues that are mentioned below [12]:

- Low maturity level in IT department and software engineering;
- Management and employees are typically overwhelmed with the routine business, leaving some space for strategic problems such as the process and quality improvement;
- There is enormous demand for the knowledge transfer to the simple problems and "how to resolve it";
- CEOs and management are habitually not used to co-operating with the outsource consultants.

The world-age has been directed to the software development with new advancements. However, there is a broad range of engineers who have acquainted the different ideas to resolve different complexities, so to develop new software that is better in reliability and supportability. Numerous reports have conceded that various industries and organizations have many opportunities to capitulate enhancements on the elements relating to the consumer satisfaction, efficiency, item quality as well as TTM (Time to market) consequences by using product line advancement ideas. Another matter of interest is the demand imposed on organizations to present/establish improved functionality and items quickly in an attempt to fulfill the market needs. These objectives are exceptionally hard to meet the capacity of making individual systems.

The arrangement of software segment is a structure of linking the concepts involving: algorithms, information sets, invocations of the tasks and the relations between different data sets. The central way of this idea is hypothetical in such a manner that conjectural basic functions have a similar impact under the various depictions. However, this research describes the points of interest and clarifications that are exceptionally precise. Testing and designing of the core structure of software development is the most complex function. Programmer's still experienced syntax errors; however, as a comparison with some grammatical errors found in a maximum number of developed systems, it is not a central issue. Another critical issue we usually face in software development is learning only by studying at institutes vs. learning by practice at work [13] [14]. Students should be aware of all kind of practical work relevant to their field because they have to use all the possible tools in future [15].

According to the industry analysts' estimation, there are more than 225,000 software applications accessible at different marketplaces. These Applications are available for different kinds of devices. Wasserman has already conducted a survey for identifying the software engineering issues for the mobile applications in 2010. An important goal of this survey was to advance a well understanding of the engineering practices for the different mobile applications. Some important points of this survey are following [26]:

1) Most of the mobile applications demands big and more experienced team of developers to complete the application before time with all the required functionalities;

2) There was a shrill divide between the "native" software applications, which run completely on the mobile device, and web applications with execution on the remote server;

3) Developers obeyed fairly well to commended sets of "best practices" but rarely used any formal development processes;

4) Developers did slightly organized chasing of their development struggles and collected few metrics.

Wasserman described the problems of application development in a good manner in his survey. Different issues had been highlighted in the survey but not described the level and chances of these problems in any scenario. In our research work we have identified several software engineering issues and also highlighted the level and occurrence chances of these problems.

## 3. Difficulties in Software Engineering

The complexity that occurs by mistake typically refers to the matters that are formed as result of the interaction, which may be addressed by using proper explanation; for example writing the facts and rolling the assembly code or delays in batch processing systems, etc. Alternatively, essential problems deal with matters that are inclined by a preexisting condition that desires to be determined, but discovering best solution looks rather problematic, and a user must employ all radical solutions by use of this program particularly.

Different studies have discussed that influences corresponding with the unintentional complexity usually require some time for alterations, but many developers these days typically spend lots of time addressing substances relating to the essential complexity [16]. Now, pay our attention to natural possessions regarding the complicated matter of progressive software schemes on issues regarding complexity, invisibility, changeability, conformity, time constraints, budget constraints, and security. Software engineering issues are shown in Figure 2.

**Invisibility:** Softwares are very invisible and cannot sometimes be visualized. Geometric conceptions are potent implements. The framework of development aids both, designer and client to evaluate the spaces and flow of traffic. Geometric accuracy is précised in the geometric notion. The sanity of software is not certainly embedded in the area; hence, there is not any geometric depiction prepared related to how they map for the land coverage, figures for connectivity representations for the computers. The graphs frequently represent the dependency patterns, a surge of control, name-space connections and data flow [17]. Moreover, charts seem to be fewer classified and of non-planar practices. A noble way to launch projected control over the same construction is by way of imposing link cutting approach between hierarchical graph type structures. Despite the development in restricting and simplifying the software structures, certainly they will hold their invisible practice; therefore some vital speculative tools are not utilized strategically. This difficulty will not only cause to delay scheme procedure within one's observance, but it may also hinder the mind considerations rigorously.

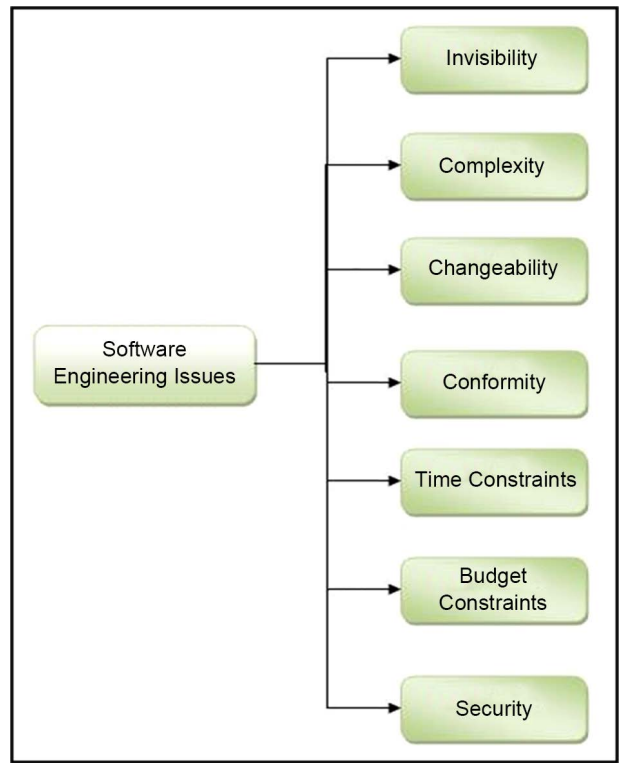

**Figure 2.** Software engineering issues.

**Complexity:** Some software parts involve a high level of difficulty regarding their scope, compared to various other developments, since all are dissimilar in some ways. If there have resemblances between two modules, they are imitated into an opened or closed. Also, software schemes are at change overwhelming in contrast to automobiles, buildings, and computers since repeated components may thrive under these situations. Digital computer projects may pose an excessive complexity as compared to the certain developments. Software projects comprise of new verdicts in contrast to the computers. In the same way, a software product rescaling is not only a recurrence of the same essentials in increased capacities; it essentially is a growth in a diversity of rehabilitated elements. Frequently, components correlate with each other in a very nonlinear way that causes the increase in the level of exertion in contrast to linearity. Software product complexity is very vigorous property, rather than the accidental occurrence.

Lots of common issues relating to the software engineering are derived from the nonlinear scope enlargement and crucial complexity practiced during the real-time development. Due to complexity experienced, communication problems may increase amongst different team members, which may result in the plan delays and cost overruns. Moreover, other circumstances influenced by the complexity include the reduced understandability, unreliability and enumeration problems. Moreover, it will be problematic to appeal various functions, which then makes it hard to employ the programs [8].

**Changeability:** As a software of a system embodies its core functionality which needed to change with the passage of time, so as to enhance quality, reliability, efficiency, etc. It is easier to upgrade a software by making substantial changes than developing new software. Software that is efficacious involve in improving the process, and there are essentially two various processes involved during this kind of change. These methods are mentioned below:

- First, deals with the effectiveness and usability of software. People use application or software systems for various causes, and certain important points may need modifications. Lots of users may feel better with basic functionalities, while others may prefer the progressive approaches to satisfy their needs.
- The second aspect relates to the software that has prospered through the lifetime of the device after its initial integration.

**Conformity:** Software users usually do not only experience issues concerning complexity while developing new software. The software engineer needs to control the arbitrary difficulty, which is superior of a primary basis sustained by systems as well as human organizations that necessitate the conformed boundaries. These boundaries will vary with the passage of time, as these are human-made creations, instead of the natural creations. In more occurrences, conformation is essential to establish its competence to conform to convinced boundaries, which usually does not have a streamlined state [18].

**Time Constraints:** Timing is a big issue that software engineers mostly face while developing the software. Developers face lots of problems due to this issue. They have strict time constraints and have to generate lots of functionalities. Due to time constraints, some functional and non-functional features remain problematic. Testing of the software is also not maturely performed due to the short time duration [19]. Time problems sometimes occur when requirements are not clearly described by the clients. Developers do have to develop the product again after getting new requirements.

**Budget Constraints:** Budget is also a big and very common issue that software engineers face. Proper cost estimation is also a challenging task [20] [21]. Clients want lots of functionalities in the very limited budget. Time and budget are directly proportional to each other [22]. Short budget projects must have to complete in a short period. If they exceed, then it considered loss for the devel-opers/software engineers.

**Security:** Security is also an important factor that the software engineers usually face. They do have to develop more secure applications and software so that no one can exploit their resources [23]. Due to the recent threats, developers need to make their software fully secure in all aspects. Software safety is an idea of developing software so that it remains to functional properly even under the malicious attack [24] [25].

## 4. Survey

We have formulated various criteria based questionnaires for the evaluation of our research and to provide an insight of what professional approaches are being adopted by various software organizations, to see if they conform to any software practices. We have conducted various research surveys by visiting different professional software development teams and took feedback from the professional software engineers to analyze the real-time problems that they are facing during the development of software applications.

We have performed the evaluation of various factors among different quality criteria that are (Projection of software development issues during S.E, Efforts required to handle such issues during S.E, Occurrence chances of such issues during S.E, Effects of such issues on software quality) and depicted them in the form of pie charts.

**Projection of Software development Issues during S.E:**

**Analysis:** Figure 3 shows the results of a survey on the projection of software development issues during software engineering. According to the results, complexity issues hold the highest value (22%) while security issues hold the least value (10%).

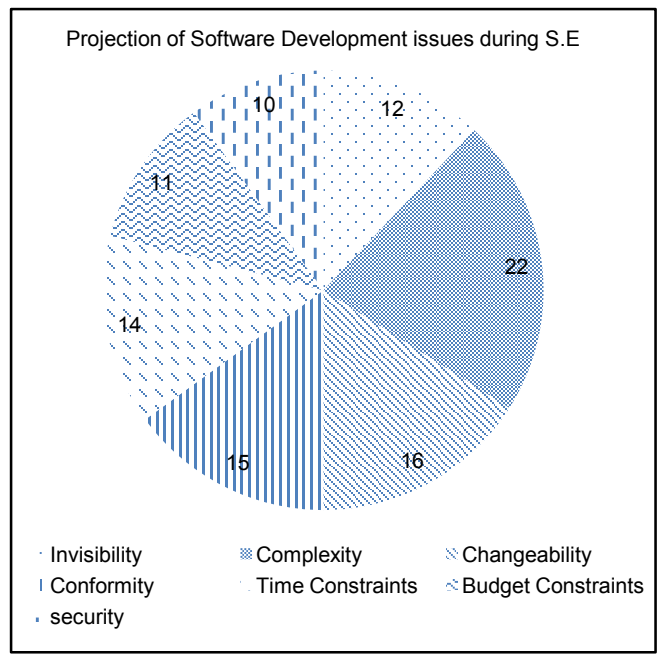

Figure 3. Projection of software development issues during S.E.

**Efforts required while handling such Issues during S.E:**

**Analysis:** According to the results shown in Figure 4, about 23% efforts are required to handle the complexity and 9% efforts required to handle invisibility issues. Complexity requires highest efforts, and invisibility requires the smallest amount of efforts.

**Occurrence chances such Issues during S.E:**

**Analysis:** Figure 5 showed that occurrence chances of complexity based problems got the highest percentile 19% and invisibility got the least percentile 11%.

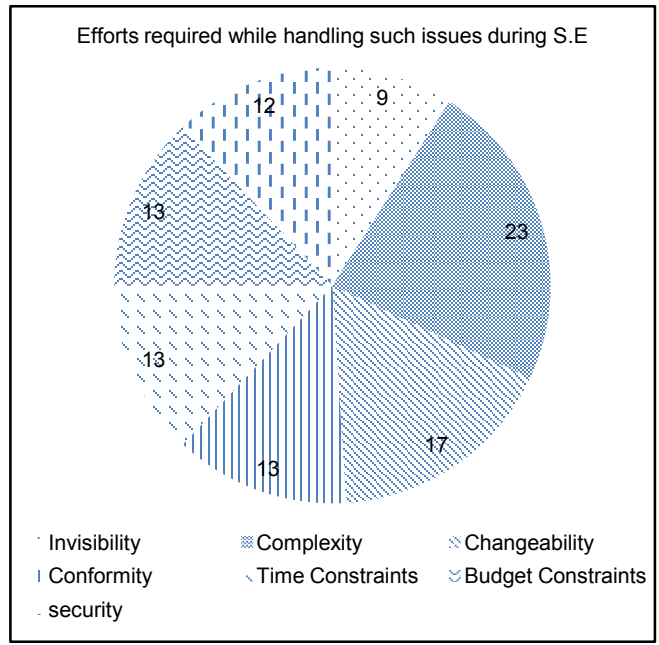

Figure 4. Efforts needed while dealing with such issues during S.E.

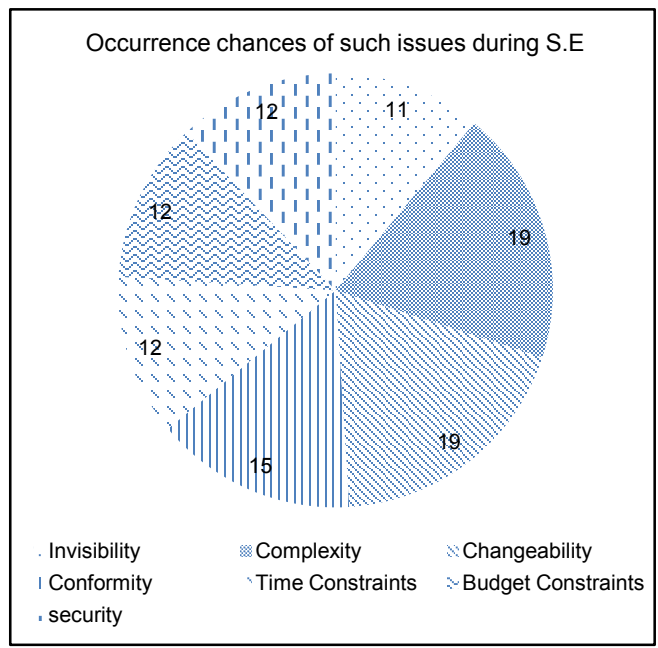

Figure 5. Occurrence chances of such issues during S.E.

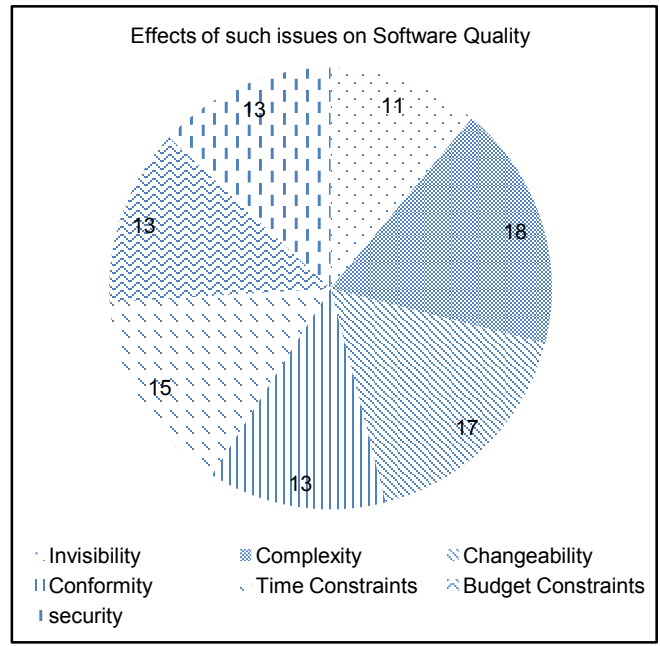

**Figure 6.** Effects of such issues on software quality.

**Effects of such Issues on Software Quality:**

**Analysis:** Results in Figure 6 showing that complexity is an issue that highly affects the software quality during software development process. Invisibility rarely affects the software quality according to the survey.

## 5. Conclusions

The main objective of this research is to see the impact of various real-time issues that most software engineers face while developing the new software. The survey is conducted by visiting different software development organizations to see the real-time issues. The survey is done on seven issues (complexity, changeability, conformity, time constraints, budget constraints, security, and invisibility) to see the percentile ratio of their occurring chances, impact, effects on software quality and efforts required to handle such issues.

Survey results showed that complexity ratio is higher in projection of S.E issues, occurring chances of these issues, efforts required to handle these issue, and effects of such issues on the software quality. In the projection of software engineering issues, security got the smallest ratio. Invisibility got the least ratio in efforts required to handle such issues, occurring probabilities, and effects of such issues on software quality according to the survey.